\documentclass{article}

\def\ROSAT{{\em ROSAT}}
\def\Einstein{{\em Einstein}}
\def\***#1{\textsc{***#1***}}

\def\putfigure#1%
{\hskip0.03\textwidth%
\hfill\makebox[0pt]{\includegraphics[width=0.54\textwidth]{#1}}\hfill}

\usepackage{graphicx}
\usepackage{emulateapj,timesfonts}

\begin{document}

\submitted{ApJ Letters, in press}

\lefthead{SEARCH FOR COUNTERPARTS OF GAMMA-RAY BURSTS}
\righthead{VIKHLININ}

\title{A search for x-ray counterparts of gamma-ray bursts with the {\em
ROSAT\/} PSPC}
\author{Alexey Vikhlinin}
\affil{Harvard-Smithsonian Center for Astrophysics, 60 Garden St.,
Cambridge, MA 02138; avikhlinin@cfa.harvard.edu}

\begin{abstract}
  We search for faint X-ray bursts with duration 10--300~seconds in the {\em
    ROSAT\/} PSPC pointed observations with a total exposure of
  $1.6\times10^7\,$s. We do not detect any events shorter than $\sim 100$~s,
  i.e.\ those that could be related to the classic gamma-ray bursts (GRB).
  At the same time, we detect a number of long flares with durations of
  several hundred seconds. Most, but not all, of the long flares are
  associated with stars. If even a small number of those long flares, that
  cannot identified with stars, are X-ray afterglows of GRB, the number of
  X-ray afterglows greatly exceeds the number of BATSE GRB. This would imply
  that the beaming factor of gamma-rays from the burst should be $>100$.
  The non-detection of any short bursts in our data constrains the GRB
  counts at the fluences 1--2.5 orders of magnitude below the BATSE limit.
  The constrained burst counts are consistent with the extrapolation of the
  BATSE $\log N - \log S$ relation. Finally, our results do not confirm a
  reality of short X-ray flashes found in the \Einstein\/ IPC data by
  Gotthelf, Hamilton and Helfand.

\end{abstract}


\section{Introduction}

Data accumulated over several years of observations by {\em ROSAT\/} and
\Einstein\/ may contain X-ray counterparts of gamma-ray bursts (GRB). Both
instruments have imaging capabilities and can provide precise positions
leading to GRB identifications. Soft X-ray data can be used to constrain the
$\log N - \log S$ distribution of gamma-ray bursts at fluxes below those
accessible by BATSE. A possibility to find X-ray afterglows of gamma-ray
bursts, similar to those recently discovered by BeppoSAX (Costa et al.\ 
1997, Piro et al.\ 1998), is another motivation for a search for soft X-ray
flares.

Gotthelf, Hamilton and Helfand (1996, GHH hereafter) reported a detection of
42 faint X-ray flashes in the \Einstein\/ IPC data. Those flashes have
typical fluxes of $10^{-10}-10^{-9}\,$ergs~cm$^{-2}$ in the 0.2--3.5 keV
energy band and a typical duration of $\lesssim 10$~s. Their detection rate
corresponds to about million per year per whole sky. Flashes are not
correlated with any known sources, including nearby galaxies.

We have performed a search for faint X-ray bursts in the pointed \ROSAT\/
PSPC observations. Using the imaging capabilities of \ROSAT, we detect
bursts as spatially localized spikes in the count rate with duration less
than about 100 seconds. Our dataset covers 2.5 times larger area and has
1.1~times longer exposure than the GHH \Einstein\/ IPC dataset.

\section{{\em ROSAT\/} Data and analysis}

\ROSAT\/ mirrors focus soft X-rays in the 0.1--2.4~keV energy band. The
prime detector, PSPC, covers a circular region with an area of 2.7~deg$^2$.
The mirror vignetting is approximately a parabolic function of the off-axis
angle; it drops to 50\% at the edge of the field of view (FOV). The angular
resolution varies from $\sim 30\arcsec$ (FWHM) on-axis to $\sim 2\arcmin$ at
the edge of the FOV. The on-axis effective area is 450 cm$^2$ at 1~keV. The
PSPC background is dominated by cosmic X-rays (Snowden et al.\ 1992). The
PSPC count rate is not saturated for source fluxes up to $\sim
1000$~phot~s$^{-1}$. These properties of the \ROSAT\/ PSPC are relevant for
our work; for a detailed description, see Tr\"umper (1983), Aschenbach
(1988) and Pfeffermann et al.\ (1987).

We used \ROSAT\/ PSPC pointed observations with exposures $>500\,$s.
Pointings to clusters of galaxies and supernova remnants were excluded
because extended emission from the target fills a significant part of the
FOV and complicates the analysis. We also excluded all pointings in the
direction of high Galactic absorption ($N_H>10^{21}\,$cm$^{-2}$), thus
omitting all Galactic plane targets. These selections leave 2256 individual
observations with the total exposure of $1.6\times 10^7\,$s.

Standard \ROSAT\/ data products have a list of good time intervals for each
observation. These intervals exclude target occultations by the Earth, the
spacecraft passages through the South Atlantic Anomaly and radiation belts,
intervals of poor aspect solution, and all detected instrument malfunctions.
As is advised in Snowden et al.\ (1994), we excluded time intervals with the
master veto rate $>170$~cnt~s$^{-1}$. The soft part of the PSPC bandpass
($E<0.4$~keV) is affected by higher background, poorer angular resolution,
and the presence of ``afterpulse'' events (Plucinsky et al.\ 1993). We,
therefore, used only the hard, 0.5--2~keV, energy band.

\subsection{Burst Detection}\label{sec:alg}

An outline of our burst detection algorithm is as follows. We sort photons
detected in the entire FOV according to their arrival time and divide the
observation into sequences of $n$ photons, advancing the sequence in steps
of $n/2$ (that is, the sequences are overlapping). In each such sequence, we
analyze the photon coordinates searching for spatial concentrations that
would correspond to flashes from point-like sources. We use sequence widths,
$n$, of 10, 15, 20, 25, 35, 45, and 60 photons, corresponding to time
intervals of 7--40~s for the average \ROSAT\/ PSPC background of $\approx
1.4$~cnt~s$^{-1}$ in the whole FOV (Fig.~\ref{fig:bgdistr}). This matches
well the duration of the classic gamma-ray bursts (e.g., Terekhov et al.\ 
1994) and the reported \Einstein\/ flashes. To find spatial concentrations
of photons within a sequence, we create an image in sky coordinates. For
each image pixel, we count photons within the 90\% power radius of the PSF,
calculated as a function of the off-axis angle (Hasinger et al.\ 1993).  If
the number of photons within this radius exceeds the preselected detection
threshold, we consider it a burst detection.

The optimal detection threshold depends on the off-axis angle because the
size of the detect cell varies. We set detection thresholds such that the
probability of a false detection in the analyzed data ($3.6\times10^7$
photons detected in the source-free regions) is small, 0.02, and that the
distribution of false detections is uniform over the FOV. The false
detection probabilities are derived from Monte-Carlo simulations of
$5\times10^9$ photons randomly distributed inside the FOV. The thresholds
vary from 4 photons for the shortest photon sequence and the inner region of
the FOV, to 11 photons for the longest photon sequence and the edge of the
FOV. If the derived thresholds are lowered by 1 photon, one would expect
$\sim 3$ false detections in the whole analyzed data.

The use the photon sequences of a fixed length rather than fixed time
intervals greatly reduces a possibility of false detections during the
intervals of high background usually caused by scattered solar X-rays.  The
count rate is not used for burst detection, therefore the high background
does not cause false detections as long as it is relatively uniform over the
detector. The sensitivity to long bursts is smaller during the high
background intervals because the photon sequences span shorter time and
hence contain a smaller fraction of the burst flux. Fortunately, high
background occurs only in a small fraction of the total exposure
(Fig.~\ref{fig:bgdistr}). The loss of sensitivity during these intervals is
fully accounted for in our calculations of the detection efficiency.

\putfigure{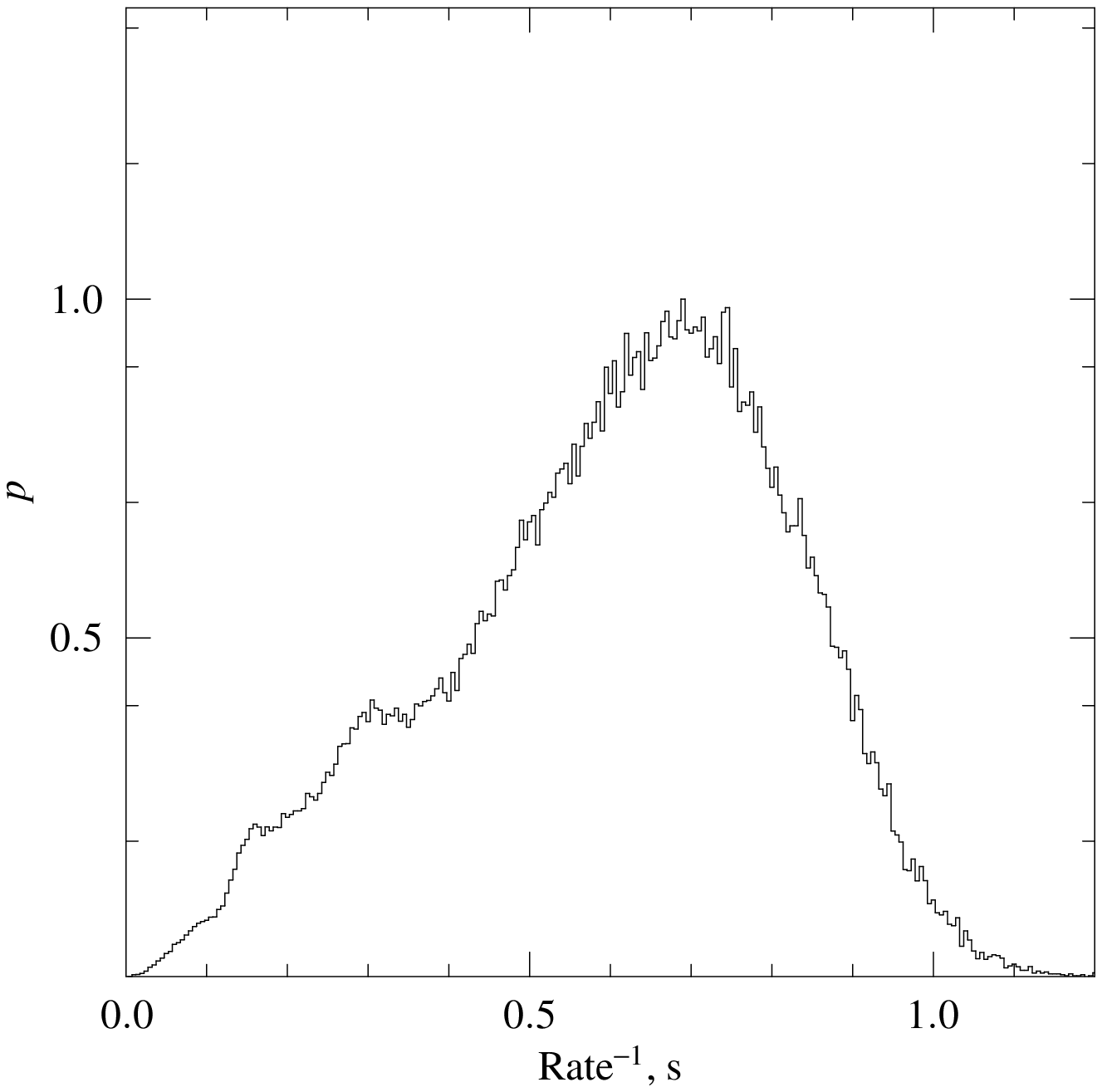} \vskip -10pt \figcaption{The distribution of the
  inverse \ROSAT\/ PSPC background rate in the whole FOV. The background
  rate was measured in sequences of 200 photons consecutively detected in
  source-free regions. The distribution represents the probability to
  measure a given count rate at a random moment in time.
    \label{fig:bgdistr} }
\medskip

Our burst detection algorithm requires absence of bright persistent sources.
For example, if there is a source with the count rate equal to that of the
background, 10 photons in each 20 photon sequence will be detected near the
same position. This is above our threshold for burst detection. Even much
fainter sources can increase the false detection probability. To avoid this
problem, we masked all sources that were detectable in the image accumulated
over the entire observation.  Sources were detected using a matched
filtering algorithm (Vikhlinin et al.\ 1995). Detection thresholds were
8--50 photons depending on the exposure and off-axis angle. Some of the
detectable sources might be bright genuine bursts. To avoid missing such
bursts, we analyzed light curves of all excluded sources.

\subsection{Detection sensitivity}

The probability to detect a burst of a given physical flux is a complex
function of the burst flux, duration, off-axis angle, and the background
intensity. It is also affected by the Poisson scatter of burst flux. The
off-axis angle determines mirror vignetting and detection thresholds.
Background rate and burst duration define a fraction of the burst flux
within the $n$-photon sequence.  To include all these effects into the burst
detection probability, we used Monte-Carlo simulations. The burst position
was simulated randomly inside the FOV. The burst flux was multiplied by the
corresponding vignetting correction. The number of burst photons was
simulated from the Poisson distribution. Photon positions were simulated
according to the local PSF. Photon arrival times were simulated assuming an
exponential light curve, $\exp\,(-t/t_0)$, with a random start time. The
background intensity was simulated from the distribution in
Fig~\ref{fig:bgdistr}.  Positions of background photons were simulated
according to the average PSPC exposure map. The burst detection algorithm
was applied to the simulated data.  Repeating the described simulations many
times, we derived the probability to detect bursts of a given flux and
duration.

Figure~\ref{fig:sens} shows the on-axis burst flux corresponding to an 80\%
detection probability as a function of burst duration $T_{90}$ --- time
interval over which 90\% of the burst flux is emitted. The limiting
sensitivity varies slowly between 11 and 18 photons for $2<T_{90}<30\,$s.
For longer bursts, a significant fraction of flux is outside the longest
photon sequence, and the limiting sensitivity correspondingly increases.

\putfigure{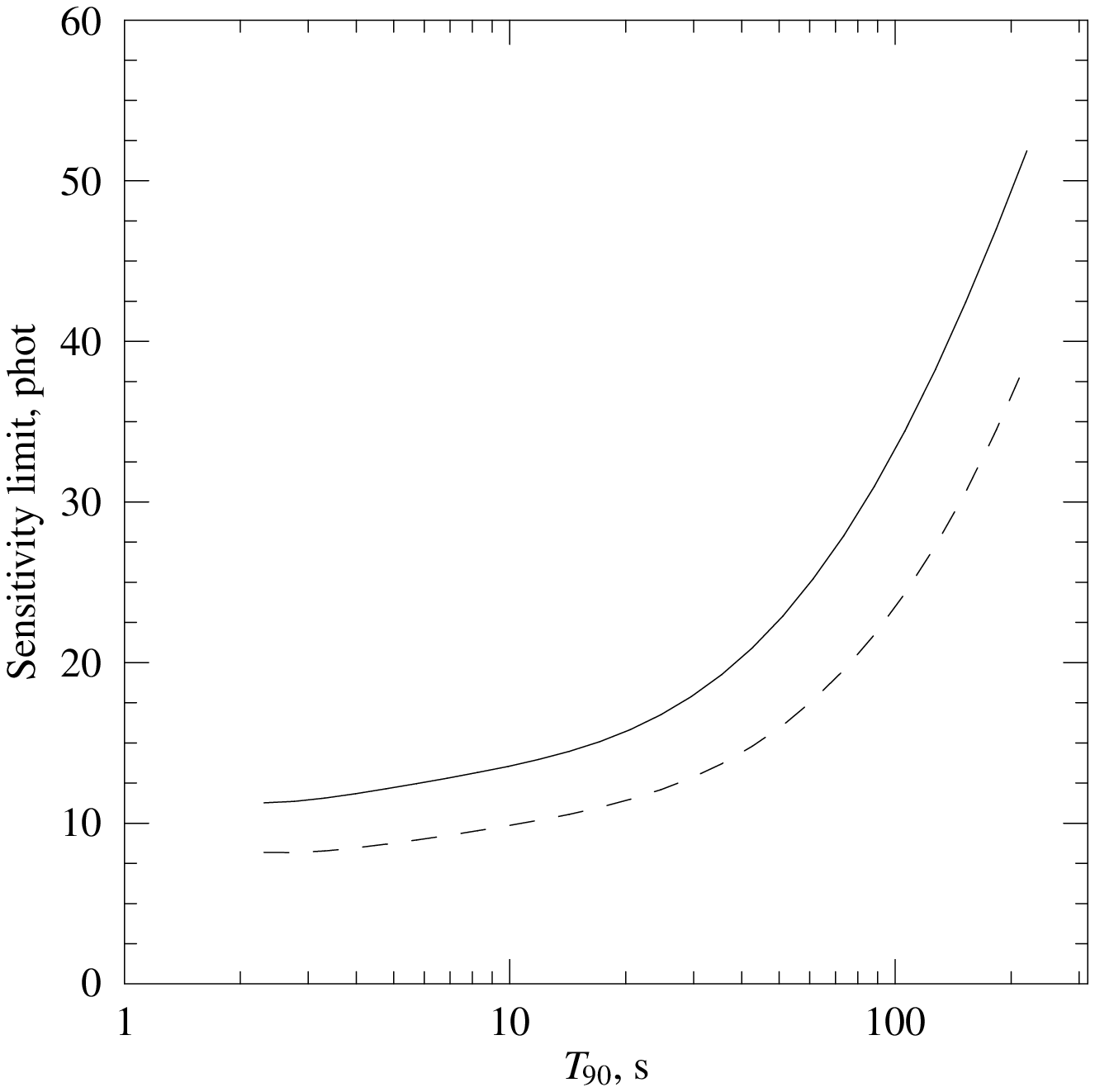} \vskip -10pt \figcaption{The on-axis burst flux
  corresponding to a 80\% detection probability as a function of burst
  duration (solid line). The dashed line shows flux at which 37\% of bursts
  are detected.
    \label{fig:sens} }
\medskip

\section{Results}

No bursts were found in the source-free regions\footnote{Earlier versions of
  the \ROSAT\/ data processing contained an error which resulted in
  assigning the same 0.5\arcsec\ detector pixel to 5--30 consecutively
  detected photons. The \ROSAT\/ data do not contain this error any longer.
  We, however, were able to detect such ``events'' in some previously
  processed datasets.}.  All bursts detected in the light curves of detected
sources were longer than several hundred seconds (see below).  Therefore,
short, 10~s, flares similar to those detected by GHH in the \Einstein\/
data, were not found in any place of the \ROSAT\/ FOV. Our results clearly
contradict to those of GHH and so a detailed comparison of the two searches
must be made.

The total exposure of our \ROSAT\/ dataset, $1.6\times 10^7\,$s, is slightly
longer than the \Einstein\/ exposure $1.5\times10^7\,$s. The geometric area
of the \ROSAT\/ PSPC FOV is 2.5~deg$^2$, while the \Einstein\/ IPC covers
only 1~deg$^2$. In both instruments, the mirror vignetting is approximately
parabolic and equals $\sim 50\%$ at the FOV edge; if anything, vignetting is
more severe for the \Einstein\/ IPC. Therefore, our \ROSAT\/ data has an
effective coverage (the product of exposure and the FOV area) 2.7 times
larger than that of \Einstein, which means that 2.7 more flares should be
found in the \ROSAT\/ data, if the two searches have a similar sensitivity.
GHH flares have $T_{90}<10$~s. Our sensitivity limit for this duration is 14
photons. The average flux of GHH flares is 11 and 7 photons for ``soft'' and
``hard'' events, respectively. Since 37 out of 42 GHH flares were detected
in the region where sensitivity is $<60\%$ of that on-axis, the average
on-axis flux is conservatively larger than 18.3 and 11.7 photons for
``soft'' and ``hard'' events, respectively. Our 0.5--2~keV energy band only
partially overlaps with the \Einstein\/ energy band 0.16--3.5~keV used in
GHH, and so conversion of count rates depends on the source spectrum. For
sources with normal spectra, the \ROSAT\/ PSPC is more sensitive.  For
example, for $N_H=10^{21}\,$cm$^{-2}$ and a typical AGN spectrum with a
photon index 1.5--2, the PIMMS software predicts the \ROSAT\/ count rate
1.3--1.5 times that of \Einstein. GHH derive a photon index of $\sim 2$ for
their hard flares, so we can conservatively assume that \ROSAT\/ count rates
should be at least 30\% higher for these events.  For ``soft'' events,
\ROSAT\/ sensitivity should be even higher.  Therefore, we expect \ROSAT\/
on-axis fluxes of at least 23.8 and 15.2 photons for soft and hard flares,
respectively. This is above our 80\% sensitivity limit
(Fig.~\ref{fig:sens}), and so at least $\sim 40$ hard and $\sim 50$ soft
flares are expected in the \ROSAT\/ data.  To conclude, the discrepancy
between \ROSAT\/ and \Einstein\/ cannot be explained by a difference in
sensitivity.

We can offer only one astrophysical explanation of this discrepancy --- a
possibility of an unusual flare spectrum causing low \ROSAT\/ fluxes. One
can estimate how unusual this spectrum should be. The 37\% sensitivity
limit, at which the sky coverage of our search becomes equal to that in GHH,
corresponds to 10 photons for $T_{90}<10\,$s. To make the \ROSAT\/ flux
below this limit, the \Einstein\/ count rate must be at least 1.2 times
higher than that of \ROSAT. This is only possible if the photon index is
very small ($<0.5$) or the spectrum is strongly self-absorbed
($N_H>10^{22}\,$cm$^{-2}$). We cannot exclude either flat spectrum or
absorption, but they seem incompatible with the flare photon index of 2
derived by GHH.

\subsection{Light curves of detectable sources}\label{sec:lcurves}

We also analyzed light curves of 93850 sources which were detected and
excluded from our main burst search. The light curves were extracted within
the 90\% power radius around the source position. Source photons were binned
into 10, 20, 40, \ldots, and 320~s time intervals. For a burst detection, we
required that the peak flux exceeded 5 photons in a time bin, that the peak
flux was at least three times higher than the average over the observation,
and that the Poisson probability of this deviation was less than
$5\times10^{-7}$ (i.e., a $5\,\sigma$ detection). These criteria correspond
to a sensitivity approximately equal to the sensitivity of the burst
detection in the source-free regions. \ROSAT\/ is routinely wobbled by $\pm
2-3\arcmin$ during the observation. The wobbling causes a spurious
variability in a small fraction of the FOV attenuated by the window support
structure of the PSPC. We excluded these regions from the source variability
analysis. We also excluded 20\% of the total exposure affected by high
background.

We detected 141 bursts in the source light curves. The integral flux in
these events was in the range 10--1000 photons. The quiescent flux was
consistent with zero in approximately half of the bursting sources. All but
one burst have $T_{90}>200\,$s. Examination of the Digitized Sky Survey
(DSS) plates has shown that 112 bursting sources were Galactic stars.  One
burst comes from the X-ray binary LMC~X-4. Dates and positions of the
remaining 28 bursts are listed in Table~\ref{tab:list}. Many of these
non-identified events still have star-like optical counterparts in the DSS,
but the faintness of the counterpart (fainter than $\sim16^{\rm m}$) makes a
reliable identification with Galactic stars impossible. Some bursting
sources are probably extragalactic. Five of 28 flares are detected in the
general direction of M31. The only short burst (8 photons in $\sim 20\,$s)
is located within the optical boundaries of M81. The peak luminosity in this
event was $\approx 7\times10^{38}\,$erg~s$^{-1}$, assuming that it was
located in M81 at $D=3.5\,$Mpc. This was $40$ times above the source
quiescent luminosity, but only a small fraction, 4.5\%, of the flux was
emitted during the burst. These properties resemble Galactic X-ray bursters.

\tabcaption{\centerline{Long bursts without a stellar identification}
\label{tab:list}}
\begin{center}
\footnotesize
\begin{tabular}{cccl}
\hline
\hline
\multicolumn{1}{c}{Date}     & R.A.       & Decl.   & \multicolumn{1}{c}{ID}\\
(dd/mm/yy)                   & (J2000)    & (J2000) & \\
\hline
05/07/91 & 00 31 26.5 & $+$26 41 38 & star-like\\
25/07/92 & 00 38 25.3 & $+$31 23 50 & uncertain\\
27/07/91 & 00 41 23.0 & $+$41 49 08 & uncertain\\
26/07/91 & 00 44 57.4 & $+$41 59 25 & star-like\\
10/01/93 & 00 45 31.4 & $+$41 54 24 & (a)      \\
31/01/93 & 00 47 32.8 & $+$42 50 03 & star-like\\
25/07/91 & 00 49 47.9 & $+$42 20 36 & star-like\\
19/07/90 & 03 34 45.3 & $-$25 37 27 & star-like\\
27/03/92 & 05 28 11.8 & $-$12 39 15 & star-like\\
29/09/93 & 06 00 35.4 & $-$38 50 25 & uncertain\\
25/08/93 & 06 45 18.8 & $+$82 48 23 & star-like\\
22/08/93 & 07 14 36.3 & $+$85 47 46 & star-like\\
03/04/91 & 07 33 26.6 & $+$31 44 08 & star-like\\
29/09/93 & 09 55 26.5 & $+$69 09 48 & in M81\\
18/05/92 & 09 56 01.7 & $-$05 12 41 & star-like\\
11/05/93 & 10 34 09.4 & $+$22 58 45 & star-like\\
20/12/91 & 11 19 03.2 & $+$07 49 27 & star-like\\
30/12/91 & 12 35 30.0 & $+$00 30 51 & star-like\\
21/12/91 & 12 49 00.3 & $-$06 28 09 & star-like\\
03/02/93 & 13 38 15.1 & $-$19 54 29 & star-like\\
26/11/91 & 13 39 09.1 & $+$48 21 10 & star-like\\
02/07/94 & 13 41 21.6 & $+$29 02 02 & star-like\\
23/07/90 & 14 17 22.5 & $+$25 05 16 & star-like\\
23/01/92 & 15 04 18.8 & $+$10 37 24 & nothing  \\
16/03/91 & 16 40 56.3 & $+$70 42 38 & star-like\\
07/08/93 & 16 39 51.1 & $+$70 51 34 & star-like\\
31/05/93 & 22 14 21.5 & $-$16 53 56 & uncertain\\
01/06/92 & 23 20 38.8 & $+$08 21 09 & star-like\\
\hline
\end{tabular}
\smallskip

$^{\rm a}$ --- supersoft source in M31 (White et al.\ 1995)
\end{center}
\medskip

A high quiescent flux of the short burst and long time scales of all other
bursts indicate that the detected events are not related to X-ray flares
detected by \Einstein.

\section{Discussion}

\label{sec:lognlogs}

Ginga and BeppoSAX observations show that classic gamma-ray bursts are
accompanied by a powerful, rapidly varying emission in the X-ray band
(Yoshida et al.\ 1989, Frontera et al.\ 1998). We would detect this emission
if the GRB had fallen inside the \ROSAT\/ FOV. Therefore, the non-detection
of any short events in our \ROSAT\/ can be used to constrain the burst
counts at a flux limit below that accessible by BATSE. A non-detection of
any events in our search translates to a 90\% upper limit of $7.2\times10^4$
bursts per year in the whole sky. Our limiting fluence is $\sim 20$ photons
for $T_{90} \sim 10-100\,$s. This corresponds to
$2.6\times10^{-10}\,$erg~cm$^{-2}$ in the 0.5--2~keV band, conservatively
assuming the maximum absorption $N_H=10^{21}\,$cm$^{-2}$ in all pointings,
and a 0.5--2~keV photon index $\Gamma=1$, consistent with the average X-ray
spectrum of Ginga bursts and the BeppoSAX spectrum of GRB~970508. The
average ratio of 50--300~keV and 0.5--2~keV fluxes of Ginga gamma-ray bursts
is $\sim 60$ (Yoshida et al.\ 1989\footnote{To calculate this number, we
  converted $1.5-10/1.5-375$~keV flux ratios provided by Yoshida et al.\ 
  into $0.5-2/50-300$~keV flux ratios assuming a thermal bremsstrahlung
  spectrum}). BeppoSAX spectrum of GRB~970508 (Frontera et al.\ 1998) yields
a $50-300/0.5-2$~keV flux ratio of $\sim 15$ for a primary, hard pulse, and
$\sim 1$ for a secondary, soft pulse. This range of observed GRB spectra
corresponds to a limiting 50--300~keV fluence of
$2.6\times10^{-10}$--$1.5\times10^{-8}\,$erg~cm$^{-2}$ for our search. This
is a 2.5--1 orders of magnitude improvement in sensitivity over BATSE
(Fig~\ref{fig:lnls}).

With the present data, we cannot exclude a possibility that some of the
detected long flares are X-ray afterglows of gamma-ray bursts, similar to
those detected by BeppoSAX. We consider this possibility unlikely, mainly
because 112 out of 141 these events are confidently identified with Galactic
stars. Of the remaining 28 events, 27 have optical counterparts in the
Digitized Sky Survey plates obtained years before the X-ray event. This is
quite different from the properties of the optical transients associated
with BeppoSAX GRB.  For example, the 970508 transient was $R=19.8$ in the
maximum, that is, below or just at the limit of the DSS sensitivity.
Therefore, flaring stars or perhaps AGN is a likely nature of the detected
long flares. If, however, just several of them are indeed GRB afterglows, an
immediate implication is that gamma-ray bursts are highly collimated,
because the long X-ray flares occur at a rate $\sim 10^5$ per year compared
to the BATSE rate of $\sim 700$ GRB per year and because in X-rays, we do
not see main bursts in the long flares.

\putfigure{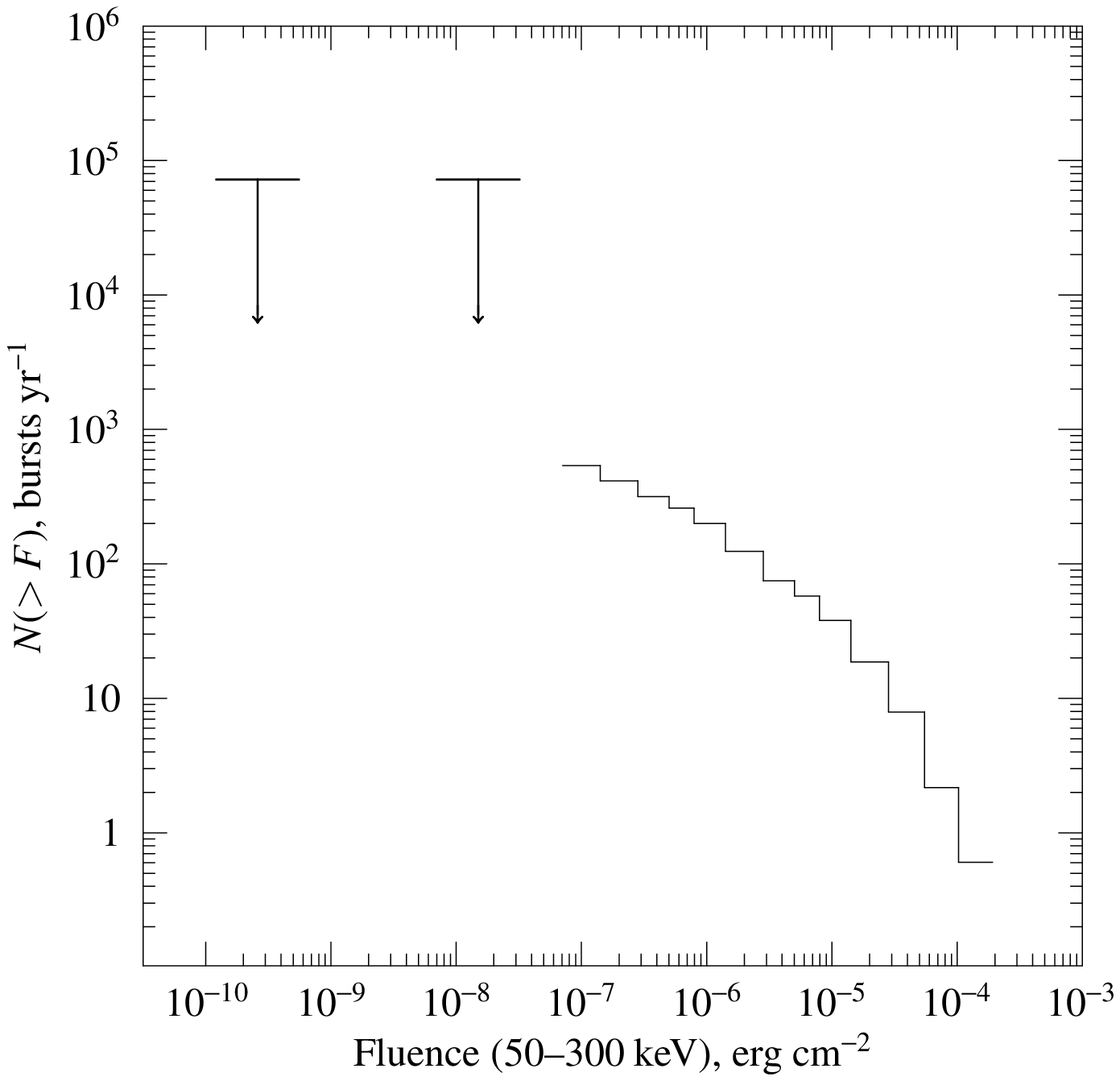} \figcaption{The $\log N - \log S$ distribution for
  gamma-ray bursts. The histogram shows BATSE burst counts from Petrosian \&
  Lee (1996). Upper limits correspond to a non-detection of soft X-ray
  bursts by \ROSAT. High and low flux points correspond to the average Ginga
  GRB spectrum and to the BeppoSAX spectrum of GRB~970508, respectively (see
  text). \label{fig:lnls}}
%

\acknowledgements

This work was supported by the Smithsonian Institution postdoctoral
fellowship and the RBRF grant 95-02-05933. The author thanks R.\ A.\ Sunyaev
for the encouragement of this work, and M.\ Markevitch, W.\ Forman, P.\
Gorenstein, and H.\ Tananbaum for interesting discussions.

\end{document}